\documentclass[sigconf]{acmart}
\AtBeginDocument{%
  \providecommand\BibTeX{{%
    \normalfont B\kern-0.5em{\scshape i\kern-0.25em b}\kern-0.8em\TeX}}}

\setcopyright{acmcopyright}
\copyrightyear{2024}
\acmYear{2024}
\setcopyright{rightsretained}
\acmConference[SIGCSE 2024]{Proceedings of the 55th ACM Technical Symposium on Computer Science Education}{March 20--23, 2024}{Portland, Oregon, USA}
\acmBooktitle{Proceedings of the 55th ACM Technical Symposium on Computer Science Education (SIGCSE 2024), March 20--23, 2024, Portland, Oregon, USA}
\acmDOI{}
\acmISBN{}
\begin{document}

%%
%% The "title" command has an optional parameter,
%% allowing the author to define a "short title" to be used in page headers.
\title{Developing a Tool to Automate Extensions to Support a Flexible Extension Policy}

%% Tools to Automate Extensions to Support a Flexible Extension Policy 
%% Automated Tools to Support Flexible Extensions for Student Success and Well-Being

%%
%% The "author" command and its associated commands are used to define
%% the authors and their affiliations.
%% Of note is the shared affiliation of the first two authors, and the
%% "authornote" and "authornotemark" commands
%% used to denote shared contribution to the research.

\author{Jordan Schwartz}
\orcid{0009-0002-0691-3815}
\affiliation{%
  \institution{University of California Berkeley}
  \city{Berkeley}
  \state{CA}
  \country{USA}
}
\email{jordanschwartz@berkeley.edu}

\author{Madison Bohannan}
\orcid{0009-0007-4131-9551}
\affiliation{%
  \institution{Massachusetts Institute of Technology}
  \city{Cambridge}
  \state{MA}
  \country{USA}
}
\email{mkcb@berkeley.edu}

\author{Jacob Yim}
\orcid{0000-0002-6174-2644}
\affiliation{%
  \institution{University of California Berkeley}
  \city{Berkeley}
  \state{CA}
  \country{USA}
}
\email{jacobyim@berkeley.edu}

\author{Yuerou Tang}
\orcid{0000-0003-2574-392X}
\affiliation{%
  \institution{University of California Berkeley}
  \city{Berkeley}
  \state{CA}
  \country{USA}
}
\email{yuerou.tang@berkeley.edu}

\author{Dana Benedicto}
\orcid{0009-0000-9812-6127}
\affiliation{%
  \institution{University of California Berkeley}
  \city{Berkeley}
  \state{CA}
  \country{USA}
}
\email{dbenedicto@berkeley.edu}

\author{Charisse Liu}
\orcid{0009-0008-5537-8891}
\affiliation{%
  \institution{University of California Berkeley}
  \city{Berkeley}
  \state{CA}
  \country{USA}
}
\email{charisseliu@berkeley.edu}

\author{Armando Fox}
\orcid{0000-0002-6096-4931}
\affiliation{%
  \institution{University of California Berkeley}
  \city{Berkeley}
  \state{CA}
  \country{USA}
}
\email{fox@berkeley.edu}

\author{Lisa Yan}
\orcid{0009-0007-2310-3060}
\affiliation{%
  \institution{University of California Berkeley}
  \city{Berkeley}
  \state{CA}
  \country{USA}
}
\email{yanlisa@berkeley.edu}

\author{Narges Norouzi}
\orcid{0000-0001-9861-7540}
\affiliation{%
  \institution{University of California Berkeley}
  \city{Berkeley}
  \state{CA}
  \country{USA}
}
\email{norouzi@berkeley.edu}

%%
%% By default, the full list of authors will be used in the page
%% headers. Often, this list is too long, and will overlap
%% other information printed in the page headers. This command allows
%% the author to define a more concise list
%% of authors' names for this purpose.
%%\renewcommand{\shortauthors}{Trovato and Tobin, et al.}

%%
%% The abstract is a short summary of the work to be presented in the
%% article.
\begin{abstract}
In this work, we present the development of an automated extension tool to assist educators and increase the success and well-being of students by implementing flexible extension policies. Flexible extension policies materialize in many ways, yet there are similarities in students' interactions with them; students tend to request multi-day long extensions repeatedly. In courses with hundreds or potentially thousands of students, providing a system to support this extension request demand is not possible given most currently available resources and limited staff. As such, a tool is necessary to help automate flexible extension processes. The development of this tool should reduce staff load while increasing individualized student support, which can be used in varying ways for different extension policies. 

Our research questions are: \textbf{RQ1:} Does the extension tool reduce barriers and stigma around asking for assistance? \textbf{RQ2:} Does the tool lessen the wait time between requesting and receiving an extension, and how does the tool improve students' learning experience in the course? These questions will help inform us about how an automated tool for flexible extensions helps support growing course sizes and students who may not otherwise receive the support they need for their success and well-being in the course.
\end{abstract}

%%
%% The code below is generated by the tool at http://dl.acm.org/ccs.cfm.
%% Please copy and paste the code instead of the example below.
%%
% \begin{CCSXML}
% <ccs2012>
%  <concept>
%   <concept_id>00000000.0000000.0000000</concept_id>
%   <concept_desc>Do Not Use This Code, Generate the Correct Terms for Your Paper</concept_desc>
%   <concept_significance>500</concept_significance>
%  </concept>
%  <concept>
%   <concept_id>00000000.00000000.00000000</concept_id>
%   <concept_desc>Do Not Use This Code, Generate the Correct Terms for Your Paper</concept_desc>
%   <concept_significance>300</concept_significance>
%  </concept>
%  <concept>
%   <concept_id>00000000.00000000.00000000</concept_id>
%   <concept_desc>Do Not Use This Code, Generate the Correct Terms for Your Paper</concept_desc>
%   <concept_significance>100</concept_significance>
%  </concept>
%  <concept>
%   <concept_id>00000000.00000000.00000000</concept_id>
%   <concept_desc>Do Not Use This Code, Generate the Correct Terms for Your Paper</concept_desc>
%   <concept_significance>100</concept_significance>
%  </concept>
% </ccs2012>
% \end{CCSXML}

% \ccsdesc[500]{Do Not Use This Code~Generate the Correct Terms for Your Paper}
% \ccsdesc[300]{Do Not Use This Code~Generate the Correct Terms for Your Paper}
% \ccsdesc{Do Not Use This Code~Generate the Correct Terms for Your Paper}
% \ccsdesc[100]{Do Not Use This Code~Generate the Correct Terms for Your Paper}

\renewcommand{\shortauthors}{Schwarz et al.}
% %%
%% Keywords. The author(s) should pick words that accurately describe
%% the work being presented. Separate the keywords with commas.

%% A "teaser" image appears between the author and affiliation
%% information and the body of the document, and typically spans the
%% page.
% \begin{teaserfigure}
%   \includegraphics[width=\textwidth]{sampleteaser}
%   \caption{Seattle Mariners at Spring Training, 2010.}
%   \Description{Enjoying the baseball game from the third-base
%   seats. Ichiro Suzuki preparing to bat.}
%   \label{fig:teaser}
% \end{teaserfigure}

% \received{20 February 2007}
% \received[revised]{12 March 2009}
% \received[accepted]{5 June 2009}

%%
%% This command processes the author and affiliation and title
%% information and builds the first part of the formatted document.
\maketitle

\section{Introduction}
In response to the growing size of CS courses in colleges and universities \cite{laliberte2023data}, the conventional approach of rigid due dates no longer meets the diverse learning needs of students \cite{garcia2022achieving, berns2021grading}. The rise in mental health issues and global crises has increased the number of students facing exceptional circumstances while pursuing higher education \cite{wang2020investigating}. However, the simultaneous surge in course enrollment and the demand for personalized support has created a significant challenge in providing the necessary assistance for students to excel academically and maintain their well-being. Professors, instructors, or teaching assistants (TAs) are handling this need for individualized support, resulting in inconsistencies in the treatment of students and their accommodations. This also places an emotional burden on course staff as they hear about students' often traumatic experiences. Furthermore, in larger courses, students are increasingly expected to advocate for themselves, in addition to coping with their existing challenges.

Courses have had to adapt to accommodate the growing number of students requiring support. In the past, many courses employed a system of slip days, granting each student a fixed number of days to submit assignments late throughout the course without exceptions. This system led to decision fatigue and stress for students, who had to weigh whether to use their slip days for a minor assignment or save them for a more significant task. However, there has been a recent shift towards a more individualized and equitable approach to extensions \cite{berns2021grading, chen2023value}. This involves the creation of an extensions tool that standardizes and automates the process of requesting extensions, determines who should receive them, sends out emails, and handles extension processing within Gradescope.

\section{Overview}
This extensions tool consists of a Google form to field student requests, a Google sheet to process extension requests, and an automation that can email students their extension request's status and, if applicable, create an extension for the corresponding assignment(s) in Gradescope. Many courses additionally publicize clearly which staff members have access to the data to keep the level of privacy as transparent with students as possible.

\subsection{Google Form For Extension Requests}
A Google form is made available to students alongside the release of the first assignment and remains open when extensions can be requested. This form is how students request an extension. The default questions in the form have been tailored to fit the needs of the flexible extension policies enacted so far but are editable and interchangeable as needed per course. These questions have also been honed to reduce the load on students and require minimal effort and time to request help without sharing detailed accounts of their personal situations.

\begin{table}
    \centering
\caption{Default Questions}
\label{tab:my_label}
    \begin{tabular}{|l|} \hline 
        What is your student ID?\\ \hline 
        What is your name?\\ \hline 
        Are you registered with the disabled students program?\\ \hline 
        Which assignment would you like an extension on? \\ \hline 
        How many days would you like an extension for?\\ \hline 
        Why do you need this extension? \\ \hline 
        Are you working with a partner? \\ \hline 
        What is your partner's email and student ID?\\ \hline
    \end{tabular}

\end{table}

\subsection{Google Sheet for Extension Approvals}
The Google Sheet back-end serves as the primary interface for course staff once the workflow (Google Form plus automation) is established. This back-end comprises five sheets that receive, parse, and interact with data. Data collected from the Google Form is parsed into a "roster" sheet, creating a row for each student. It records the extension duration per assignment, timestamp, disability services enrollment status (if provided), and the most recent reason for an extension request. Additionally, it indicates the extension request status (automatic, pending approval, or manual) and email status (automatic, pending approval, in queue, or manual). This user-friendly interface streamlines the process for staff, eliminating the need to send manual extension request emails to each student. It also conserves staff energy, allowing for the easy concealment or disregard of disability services enrollment and reason for extensions while still providing necessary accommodations to students when feasible.

\begin{figure}[h]
  \centering
  \includegraphics[width=\linewidth]{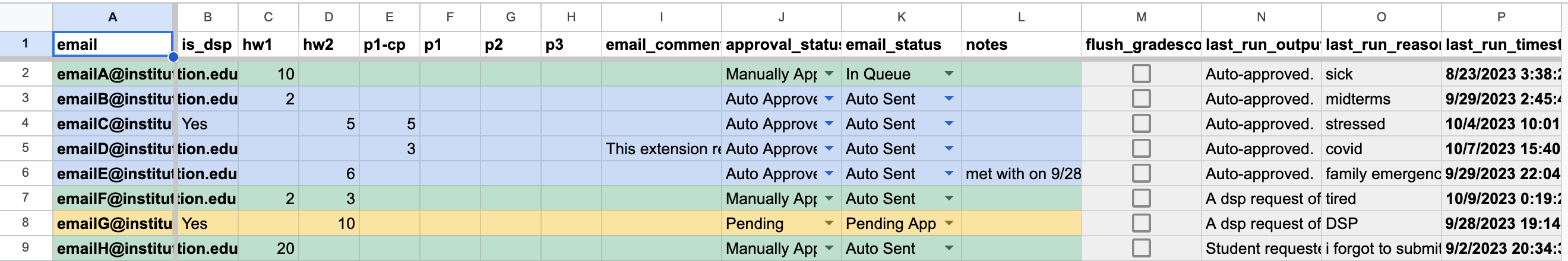}
  \caption{Example Extensions Tool Back-end}
\end{figure}

\section{Future Contributions}
The tool currently enables students to request extensions with ease and receive prompt responses. The setup involves some computer science knowledge, such as configuring a Google Sheets app script. We are currently working on a walkthrough setup video and have future plans for a more user-friendly UI that maintains flexibility while minimizing common user errors.
The tool’s adoption across numerous courses at a US public university with a selective CS program has reduced the stigma of seeking help and led to a surge in extension requests. It was originally created for a single course but expanded to meet demand for flexible extension policies. As the tool scales up to meet greater demand, we anticipate that it will need better support when errors pop up. Our current efforts are focused on an FAQ page to address common issues. In the future, we envision setting up a Google form or specific email to field requests. These requests could also be useful in updating the FAQ page.
\begin{figure}[h]
  \centering
  \includegraphics[width=\linewidth]{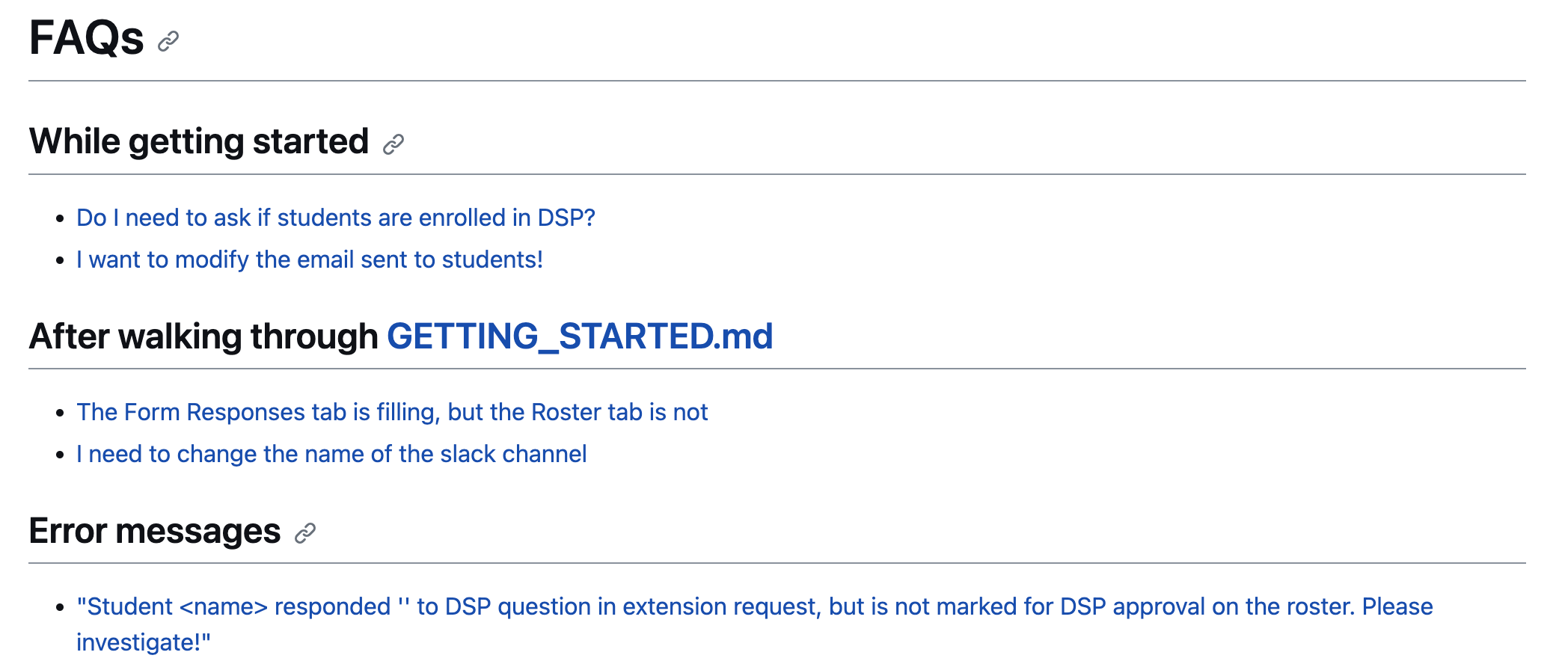}
  \caption{Example Extensions Tool Back-end}
\end{figure}

This tool was designed for one course that uses Gradescope for student submissions. As a result, it is less effective when used in courses with different assignment submission platforms. While the tool can still be used for email automation, updating assignment extensions automatically in a different grading platform is not yet possible, resulting in a time burden on course staff to do so. Our next goal is to integrate the extension request back-end with Canvas.
In its current form, the tool is explicitly tailored to the policy of the course for which it was initially created. One side effect is that the tool provides no infrastructure to reject extensions. Courses with more assignments or different policies find having to manually inform students of extension request rejections a burden. Some courses have attempted to rectify this issue by manually updating the student’s request and including a note that the request was not approved. We plan to adapt the tool to allow more flexibility in its email templates, decreasing the burden on courses with different extension policies.
Utilizing this tool in as many courses as possible will facilitate the spread of flexible extension policies and support students' success throughout their academic careers. It is our hope that this tool can meet the needs of even larger courses as class sizes continue to scale up.  

%%
%% The next two lines define the bibliography style to be used, and
%% the bibliography file.
\bibliographystyle{ACM-Reference-Format}
\bibliography{bib}

\end{document}